\documentclass{aa}  

\usepackage{graphicx}
\usepackage{comment}
\usepackage{placeins}
\usepackage{txfonts}
\usepackage[pdfpagelabels=false]{hyperref}
\hypersetup{final,colorlinks=true,linkcolor=blue,citecolor=blue,urlcolor=blue}
\newcommand{\msunyr}{\ensuremath{\mathit{M}_{\odot} {\rm yr}^{-1}}}
\usepackage{multirow}

\usepackage{tcolorbox}

\begin{document}

   \title{The dusty circumstellar environment of Betelgeuse during the Great Dimming as seen by VLTI/MATISSE \thanks{Based on observations collected at the European Southern Observatory under ESO programme 104.20V6.} \fnmsep \thanks{Based in part on observations with ISO, an ESA project with instruments funded by ESA Member States (especially the PI countries: France, Germany, the Netherlands and the United Kingdom) and with the participation of ISAS and NASA.}}

   \author{E. Cannon
          \inst{1}
          \and
          M. Montarg{\`e}s
          \inst{2}
          \and
          A. de Koter
          \inst{1,3}
          \and
          A. Matter
          \inst{4}
          \and
          J. Sanchez-Bermudez
          \inst{5,6}
          \and
          R. Norris
          \inst{7}
           \and
          C. Paladini
          \inst{8}
          \and
          L. Decin
          \inst{1,9}
          \and
          H. Sana
          \inst{1}
          \and
          J.O. Sundqvist
          \inst{1}
          \and
          E. Lagadec
          \inst{4}
          \and
          P. Kervella
          \inst{2}
          \and
          A. Chiavassa
          \inst{4}
          \and
          A. K. Dupree
          \inst{10}
          \and
          G. Perrin
          \inst{2}
          \and
          P. Scicluna
          \inst{8}
          \and
          P. Stee
          \inst{4}
          \and
          S. Kraus
          \inst{11}
          \and
          W. Danchi
          \inst{12}
          \and
          B. Lopez
          \inst{4}
          \and
          F. Millour
          \inst{4}
          \and
          J. Drevon
          \inst{4}
          \and
          P. Cruzal{\`e}bes
          \inst{4}
          \and
          P. Berio
          \inst{4}
          \and
          S. Robbe-Dubois
          \inst{4}
          \and
          A. Rosales-Guzman 
          \inst{6}
          }

   \institute{Institute of Astronomy, KU Leuven, Celestijnenlaan 200D B2401, 3001 Leuven, Belgium\\
              \email{emily.cannon@kuleuven.be}
         \and
             LESIA, Observatoire de Paris, Université PSL, CNRS, Sorbonne Université, Université de Paris, 5 place Jules Janssen, 92195 Meudon, France
        \and
        Anton Pannekoek Institute of Astronomy, University of Amsterdam, The Netherlands
        \and
        Université Côte d'Azur, Observatoire de la Côte d'Azur, CNRS, Laboratoire Lagrange, Bd de l'Observatoire, CS 34229, 06304 Nice cedex 4, France
        \and
        Max Planck Institute for Astronomy, Königstuhl 17, 69117, Heidelberg, Germany
        \and
        Instituto de Astronom\'ia, Universidad Nacional Aut\'onoma de M\'exico, Apdo. Postal 70264, Ciudad de M\'exico, 04510, M\'exico
        \and 
        Physics Department, New Mexico Institute of Mining and Technology, 801 Leroy Place, Socorro, NM 87801, USA
        \and
        European Southern Observatory, Alonso de Cordova 3107, Vitacura, Santiago, Chile
        \and
        University of Leeds, School of Chemistry, Leeds LS2 9JT, United Kingdom
        \and
        Center for Astrophysics, Harvard \& Smithsonian, 60 Garden Street, Cambridge, MA 02138, USA
        \and
        School of Physics and Astronomy, University of Exeter, Exeter EX4 4QL, UK
        \and
        NASA Goddard Space Flight Center, Exoplanets \& Stellar Astrophysics Laboratory, Code 667, Greenbelt, MD 20771, USA
             }

   \date{Received xxx,; accepted xxx}

 
 
  \abstract
   { The ‘Great Dimming’ of the prototypical red supergiant Betelgeuse, which occurred between December 2019 and April 2020, gives us unprecedented insight into the processes occurring on the stellar surface and in the inner wind of this type of star. In particular it may bring further understanding of their dust nucleation and mass loss processes.}
   {Here, we present and analyse VLTI/MATISSE observations in the N-band ($8 - 13$~$\mu$m) taken near the brightness minimum in order to assess the status of the dusty circumstellar environment.}
   {We explore the compatibility of a dust clump obscuring the star with our mid-infrared interferometric observations using continuum 3D radiative transfer modelling, and probe the effect of adding multiple clumps close to the star on the observables. We also test the viability of a large cool spot on the stellar surface without dust present in the ambient medium.}
   {Using the visibility data, we derive a uniform disk diameter of 59.02 $\pm$ 0.64 mas in the spectral range 8 to 8.75 $\mu$m. We find that both the dust clump and the cool spot models are compatible with the data. Further to this, we note that the extinction and emission of our localised dust clump in the line of sight of the star, directly compensate each other making the clump undetectable in the spectral energy distribution and visibilities. The lack of infrared brightening during the `Great Dimming' therefore does not exclude extinction due to a dust clump as one of the possible mechanisms.
   The visibilities can be reproduced by a spherical wind with dust condensing at 13 stellar radii and a dust mass-loss rate of $(2.1-4.9) \times 10^{-10}\,\msunyr$, however, in order to reproduce the complexity of the observed closure phases, additional surface features or dust clumps would be needed.}
   {}


   \maketitle
%



\section{Introduction}
\label{sec:intro}
In late 2019 / early 2020 the red supergiant (RSG), Betelgeuse ($\alpha$ Ori), experienced an unprecedented dimming event \citep{2020ATel13512....1G} with its brightness in the V-band diminishing by approximately 70\% at its minimum in February 2020. Multiple hypotheses regarding the physical processes responsible for this great drop in brightness have been proposed. These include dust nucleation above the surface obscuring the star  \citep{2020arXiv200505215S,2020RNAAS...4...39C,2020ApJ...891L..37L}, an ejection of dense chromospheric material from the southern hemisphere subsequently forming molecules and dust \citep{2020ApJ...899...68D}, photospheric cooling as a result of stochastic convective motions and pulsations \citep{ 2020ApJ...897L...9D, 2020ApJ...905...34H, 2021Alexeeva} and a `molecular plume' causing a local opacity increase \citep{2021A&A...650L..17K, 2021MNRAS.508.5757D}.   
   
To further explore and differentiate between these scenarios, spatially resolved observations of the stellar disk and direct surroundings of the star were needed. VLT/SPHERE-ZIMPOL observations from \cite{2021Natur.594..365M} provided such data and revealed that the dimming was most apparent in the southern hemisphere of the star. Based on 3D dust radiative transfer modelling and fitting of the spectral energy distribution (SED) they concluded that their data of the `Great Dimming' is in agreement with both a localised photospheric cooling and the formation of dust concentrated in the line of sight toward the southern part of the star, or the (possibly additive) effects of both phenomena. A hypothesis that is further strengthened by the recent study by \cite{2022NatAs.tmp..121T} who come to the same conclusion using photometric monitoring of Betelgeuse from the Himawari-8 meteorological satellite.


Understanding the cause of the `Great Dimming' of Betelgeuse may prove vital in unravelling the mystery behind the mechanism(s) driving copious amounts of mass from RSGs ($\dot{M} \sim 10^{-7}-10^{-5}$; e.g. \citealt{2010A&A...523A..18D, 2020MNRAS.492.5994B}). \cite{1992iesh.conf...86G} and \cite{2007A&A...469..671J} propose the mass-loss trigger is linked to surface activity, where turbulent motions, pulsations and large convective cells up-welling from the stellar envelope lower the effective gravity, potentially allowing radiation pressure to launch material. Large convective cells have been observed, detected as hot and cool spots on the stellar surface, on Betelgeuse \cite[e.g.,][]{2009A&A...508..923H,2010A&A...515A..12C,2016A&A...588A.130M,2017A&A...602L..10O} using optical and radio interferometry. \cite{2021A&A...646A.180K} pioneered a theoretical mass-loss prescription in which the wind is driven by turbulent pressure. Such turbulence may be connected to (sub-)surface activity, specifically to pulsations and the violent surfacing of large convective cells. If indeed it is possible to link surface activity to the ejection of material from the star and the consequent formation of clouds of dust it could bring us one step closer to understanding the (or at least one) wind launching mechanism of RSGs. \cite{2021MNRAS.508.5757D} These stars are the most frequent progenitors of type II-P supernov\ae{} (SNe). Recent observations \citep{2018MNRAS.476.2840M,2022ApJ...924...15J} have shown how mass loss during the RSG phase can modify the SN light curve. Close-by RSGs such as Betelgeuse are ideal candidates to search for such observational constraints due to their large angular diameters allowing for spatial resolution of the star.  

In this paper we present VLTI/MATISSE N-band observations of Betelgeuse, obtained near the V-band brightness minimum, in February 2020. We test the compatibility of the cool spot and dust clump scenarios with these data. To this end, we perform 3D radiative transfer modelling of the star and near surroundings. In Sect. \ref{sec:obs} we discuss the acquisition of the observations and the data reduction process. We then present the observational analysis in Sect. \ref{sec:analysis}. The setup and analysis of the dust radiative transfer models are described in Sect. \ref{sec:rad_trans}, Sect. \ref{sec:cool_spot} shows the same for our cool spot model. We discuss our findings in Sect. \ref{sec:discussion} and finish with a summary and conclusions in Sect. \ref{sec:summary}.


\section{Observations and data reduction}
\label{sec:obs}

MATISSE \citep{2014Msngr.157....5L, 2021arXiv211015556L} is a four-telescope beam combiner instrument at ESO's VLTI, which covers the mid-infrared range from 2.8 to 13~$\mu$m. Observations of Betelgeuse were taken with the small, medium, and large configurations of the 1.8m ATs (Auxiliary Telescopes) during three separate nights in February 2020. Each observation of Betelgeuse was preceded and/or followed by a spectro-interferometric calibrator observation. Sirius and Epsilon Leporis ($\epsilon$\,Lep) were used for that purpose.  For this study we focused on the N-band (8 - 13 $\mu$m) data, which were obtained in the high spectral resolution ($R \sim 220$) mode of MATISSE. The log of these observations is shown in Table \ref{obstab}.


The data were reduced with version 1.5.5 of the MATISSE pipeline\footnote{\url{https://www.eso.org/sci/software/pipelines/matisse/}}. In this work, we did not aim to study the accurate dust mineralogy of Betelgeuse. Hence, to optimise the SNR on our data, we thus rebinned them to a lower spectral resolution ($R \sim 30$). This step is performed in the pipeline on the fringe data before the computation of the interferometric observables.

The pipeline outputs are in the form of OIFITS files (version 2), which contain the uncalibrated interferometric observables. Six spectrally-dispersed squared visibilities, three independent closure phases (out of four in total) and four total spectra (from the four telescopes) per exposure are included.

\begin{table*}[ht]
\caption{Log of MATISSE N-band observations of Betelgeuse, where the letters A, B, and C give the sequential order of observations during a night. The column `Configuration' specifies the positions of the four telescopes.  `Pass' or `Fail' denotes whether we were able to use the observation or not.}             
\centering          
\begin{tabular}{l c c c c c }     
\hline \hline      
                     
Date & Time (UT) & Configuration & Calibrator(s) & Seeing [arcsec] & Quality Assessment  \\ 
\hline  \noalign{\vskip 2mm}    

    2020-02-08(A)  & 00:01:49 & A0-B2-D0-C1  & Sirius & 0.71 & Pass\\
    2020-02-08(B)  & 00:50:19 & A0-B2-D0-C1  & Sirius \& $\epsilon$\,Lep & 0.65  & Fail\\
    2020-02-08(C)  & 01:38:08 & A0-B2-D0-C1  & $\epsilon$\,Lep \& Sirius & 0.57  & Pass\\
    2020-02-19(A)  & 00:45:22 & K0-G2-D0-J3  & $\epsilon$\,Lep & 0.78  & Fail\\
    2020-02-19(B)  & 01:37:15 & K0-G2-D0-J3  & Sirius & 0.81  & Pass\\
    2020-02-25     & 00:41:21 & A0-G1-J2-J3  & $\epsilon$\,Lep & 1.04  & Fail\\
    \hline\noalign{\vskip 2mm}  

\end{tabular}
\label{obstab}
%
\end{table*}

   \begin{figure*}[t!]
   \centering
   \includegraphics[width=0.85\textwidth]{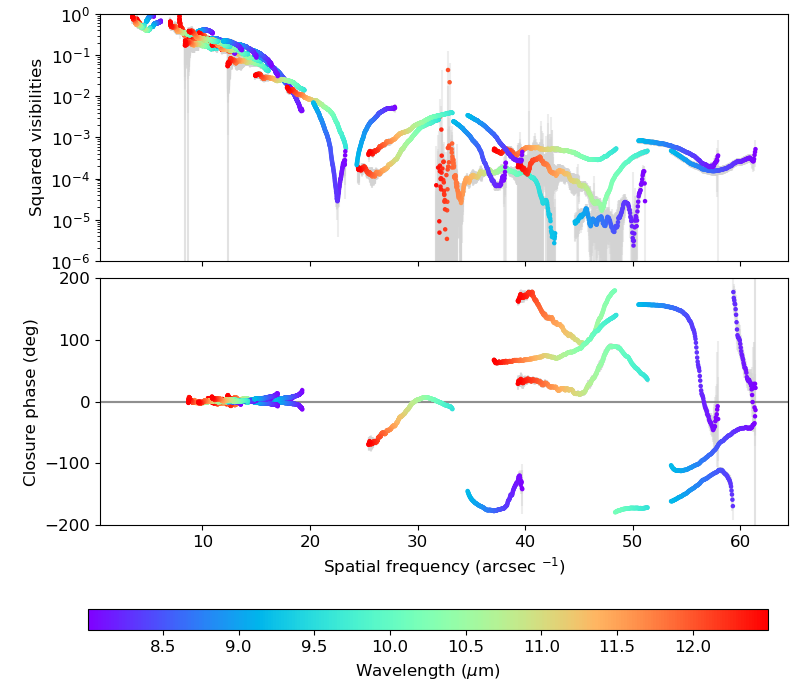}
      \caption{
      VLTI/MATISSE calibrated observations of Betelgeuse. Top: Squared visibilities with 1-$\sigma$ error bars shown in grey. Bottom: Closure phases. The colour bar indicates the wavelength. }
         \label{observations}
   \end{figure*}

To calibrate the observations, we used two calibrators, Sirius and $\epsilon$\,Lep (see Table \ref{obstab}). They were taken from the JMMC Stellar Diameter Catalog (JSDC) version 2 \citep{2016A&A...589A.112C}, with corresponding angular diameters of $6.16 \pm 0.47$ mas and $5.92 \pm 0.56$ mas, respectively. For each exposure, the visibility calibration was performed by dividing the raw squared visibility of the science target by the raw visibility of the calibrator corrected for its diameter. The calibrated total N-band spectrum of Betelgeuse was obtained on the first and third snapshots of the first night by multiplying the ratio between the target and calibrator raw fluxes measured by MATISSE at each wavelength, with a model of the absolute flux of the calibrator (Sirius). The latter was taken from the PHOENIX grid \citep[ACES-AGSS-COND;][]{2013A&A...553A...6H}. After that calibration step, we merged the four exposures to obtain a final calibrated OIFITS file for every snapshot.

After a first quality check, it appeared that some of the snapshots on Betelgeuse were corrupted and we thus excluded them from our modelling and analysis. For the first night, the visibilities of the four consecutive interferometric exposures of the second snapshot appeared very variable and differed by significantly more than $3\sigma$. A fast-varying seeing, thus implying an unstable chopping and photometry measurement, is very likely the cause of it. For the second night, the first snapshot showed nearly zero visibilities and extremely noisy closure phases, both very different from the second snapshot that showed good non-zero visibilities and closure phases. The first snapshot likely suffered from problems of fringe coherencing during the observations and then residual optical path difference (OPD) correction at the data reduction stage. Finally, for the third night where the large AT array was used, it turned out that for all the baselines but the shortest one (A0-G1), Betelgeuse was extremely resolved ($V \lesssim 10^{-3}$). That implied N-band correlated fluxes below the MATISSE N-band sensitivity limit with ATs ($\sim 5 Jy$) and thus the data were unusable.

\begin{figure}
   \centering
   \includegraphics[width=\hsize]{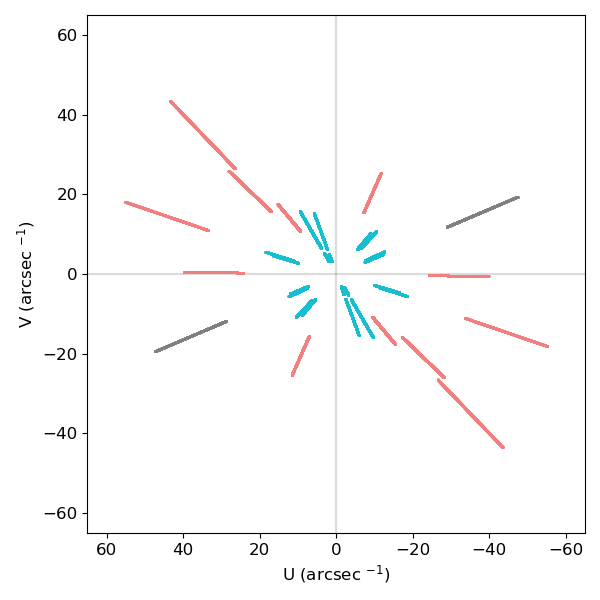}
      \caption{($u,v$) coverage of the small configuration shown in blue, medium in pink and large in grey. 
              }
         \label{UV}
\end{figure}

 Figure \ref{observations} shows the calibrated squared visibilities and closure phases used in our modelling. The error bars contain two contributions: 1) a short-term one affecting the individual 1-min exposures, as computed by the MATISSE pipeline, and 2) the standard deviation between the merged exposures. Note that the calibration error associated with the stability of the MATISSE N-band transfer function over an observing night is estimated to be of about 5\% in average seeing conditions \citep{2021arXiv211015556L} as it was the case for our night (see Table~\ref{obstab}). In our case, that was smaller than the two other error contributions. Figure~\ref{UV} shows the ($u$,$v$) coverage of the small and medium configuration, and one baseline from the large configuration. The small and medium configurations probe scales of 60~-~300~mas and 20~-~80 mas respectively. The longer baselines from the large configuration can probe even smaller scales, however, only the shortest baseline of this configuration is usable from these observations, which probes similar scales to the medium configuration. Notice the sparseness of this coverage that prevents us from performing an image reconstruction.

In order to compare the flux retrieved from MATISSE to that of the star pre-dimming we use a spectrum obtained using ISO's \citep{1996A&A...315L..27K} SWS \citep{1996A&A...315L..49D} instrument. The version of the ISO data presented in this paper correspond to the Highly Processed Data Product (HPDP) set called `A uniform database of SWS 2.4-45.4 micron spectra' by \cite{2003ApJS..147..379S}, available for public use in the ISO Data Archive\footnote{\url{http://iso.esac.esa.int/ida/}, observation ID: 69201980}.


\section{Data Analysis}
\label{sec:analysis}

\subsection{Squared visibilities and stellar diameter}
\label{sec:vis}
From the van Cittert-Zernike theorem, complex visibilities are the Fourier transform of the objects brightness distribution. A value of |V|$=1$ indicates the object is not resolved. An object is considered fully resolved when the first zero of its visibility function is observed, as is the case of the observations in this study. The squared visibilities for the three telescope configurations are shown in the top panel of Fig. \ref{observations} against spatial frequency, i.e., the length of the projected baseline divided by the wavelength. Going to larger spatial frequencies probes finer details. Throughout this paper spatial frequency is expressed in arcsec$^{-1}$
While the wavelength range of the observations extends to 13\,$\mu$m, we made a cut at 12.5\,$\mu$m due to a high noise level beyond this wavelength. The dips in the visibilities of many of the baselines in the first lobe are particularly noticeable, with a relative minimum at approximately 10\,$\mu$m. Given the wavelength these dips occur at, we suggest that this could be a result of silicate dust around the star. This is explored further with dust radiative transfer modelling in Sect. \ref{sec:rad_trans}. 
 
\begin{figure}
   \centering
   \includegraphics[width=\hsize]{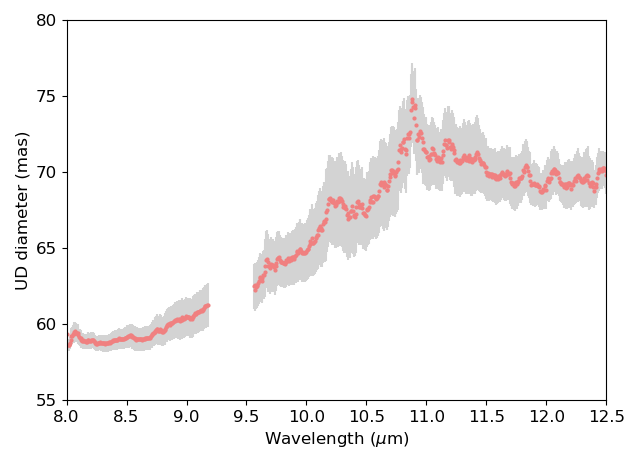}
      \caption{ Results of uniform disk fit as a function of wavelength, with 1-$\sigma$  error bars shown in grey. The wavelength region $8 - 8.75\,\mu$m suffering least from dust emission was used to fit a uniform disk diameter (see Fig.~\ref{60mas}). The gap in the 9.2 - 9.5 $\mu$m region corresponds to flagged data.
              }
         \label{UD_fit}
   \end{figure}

We performed a per spectral channel uniform disk (UD) fit to the squared visibilities obtained from the two snapshots in the small configuration (which cover much of the first lobe), the results of which can be seen in Fig.~\ref{UD_fit}. The fit is not extended towards longer baselines to avoid contamination by small scale structures.
The apparent stellar diameter appears to increase with wavelength, peaking between 10$-$11.5\,$\mu$m. The presence of dust in the circumstellar environment is likely responsible for this. We expect that, in the N-band wavelength range, we should see the least contribution from dust for wavelengths $<8.75\,\mu$m. Beyond this wavelength the contributions of dust and other components in the field of view become important, therefore, values for the disk diameter from $\lambda>8.75~\mu$m, should not be taken at face value as they are the result of a complex combination of the apparent angular diameter, resolved out flux, and the angular separation and flux ratio of components in the circumstellar environment. Previous interferometric observations of Betelgeuse where multi-component models have been applied to separate out the extended contribution at these longer wavelengths obtain disk diameters of $\approx$~53~-~57 mas at $\sim$11$~\mu$m \cite[e.g.,][]{1994AJ....107.1469D, 1996ApJ...463..336B, 2000ApJ...544.1097W,2007A&A...474..599P}. It is worth noting that differences between angular diameter measurements could be caused by different spectral bandwidths, the evolution of the circumstellar environment and in the case of resolved dust emission also the ($u,v$) coverage. 

For the purposes of this study we take the mean over the $8 - 8.75\,\mu$m interval, to obtain a disk diameter, $\theta_\star^\mathrm{MIR}$, of 59.02~$\pm$~0.64 mas.
\cite{2017A&A...602L..10O} retrieve a major axis diameter of 57.8~$\pm 0.1$\,mas from an uniform elliptical disk fit to ALMA band 7 (0.8–1.1 mm) observations. \cite{2018A&A...609A..67K} also determine the diameter in this ALMA band to be 59~$\pm 0.28$\,mas. 
VLTI/GRAVITY K-band (2.0 - 2.4 $\mu$m) data taken in the same month as our MATISSE observations give $\theta_\star^\mathrm{NIR} = 42.11 \pm 0.05$\,mas \citep{2021Natur.594..365M}, indicating that the star appears much larger in the N-band than K-band. This change in apparent diameter could be due to a close molecular layer of H$_2$O \citep{2000ApJ...538..801T, 2004A&A...421.1149O,2014A&A...572A..17M} and SiO \citep{2008A&A...480..431D, 2018A&A...609A..67K} around the star. This suggests that the diameter we measure here is a combination of the photospheric diameter and so-called MOLsphere \citep{2000ApJ...538..801T}. In reality, the opacity of these molecules varies over the N-band wavelength range, causing some variations in the apparent diameter, where SiO is most prominent under 10 $\mu$m and H$_2$0 is present across the entire range (see \citealt{2007A&A...474..599P}). While we find our N-band diameter result comparable to the disk diameter measured in the sub-mm, it is important to note that different opacity sources are dominant in each wavelength regime. For instance, H$^-$ opacity is dominant for RSGs such as Betelgeuse at sub-mm wavelengths \citep{2001ApJ...551.1073H}. 

\begin{figure}
   \centering
   \includegraphics[width=\hsize]{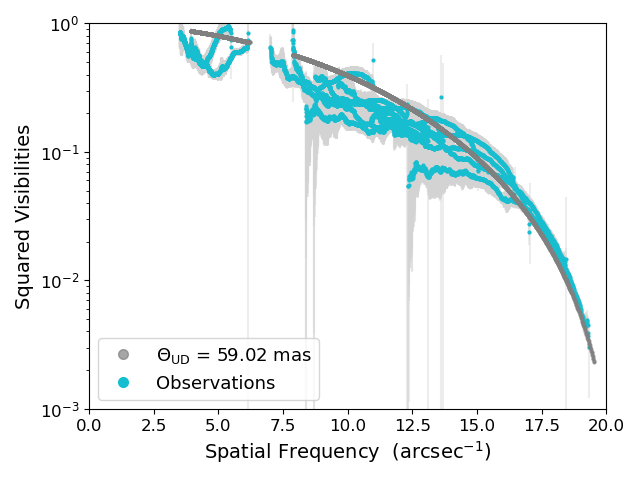}
      \caption{Squared visibilities from the small configuration are shown in blue with 1-$\sigma$ error bars shown in lightgrey. A model of a uniform disk with a diameter of 59.02 mas is plotted in dark grey. 
              }
         \label{60mas}
   \end{figure}

Since we do not have access to the angular diameter for wavelengths larger than 8.75\,$\mu$m, except by using geometrical modelling assuming a certain type of geometry, we decide to use this constant diameter value, $\theta_\star^\mathrm{MIR}$, for the photosphere. Figure~\ref{60mas} shows the first lobe part of the observations over-plotted with a 59.02 mas UD model, here, we can see that the model fits the general trend of the visibilities outside of the 10\,$\mu$m feature, which we will aim to reproduce using dust radiative transfer modelling in Sect. \ref{sec:rad_trans}.

\subsection{Closure phases}
\label{sec:cp}
Closure phases are the phase sum over 3 baselines, which allows the recovery of partial phase information that would otherwise be lost as a result of atmospheric turbulence. Deviations from 0 or $\pm 180^{\circ}$ give information on the asymmetries of the system. The closure phases are shown in the bottom panel of Fig.~\ref{observations}. The small configuration, which probes scales of 60 - 300 mas, shows little departure from 0$^{\circ}$ in the closure phases, indicating that we are not detecting significant asymmetries at these scales. However, the medium configuration, probing scales of approximately 20$-$80 mas, appears much more complex indicating deviations from a centro-symmetric system at these scales. These deviations could be caused by clumpiness in the environment around the star and/or brightness variation across the surface of the star.

An attempt to reconstruct an image was made. However, due to the sparse ($u,v$) coverage of our observations (see Fig.~\ref{UV}) we were unable to converge on a definitive solution. 

   \begin{figure*}[t]
            \includegraphics[width=\hsize]{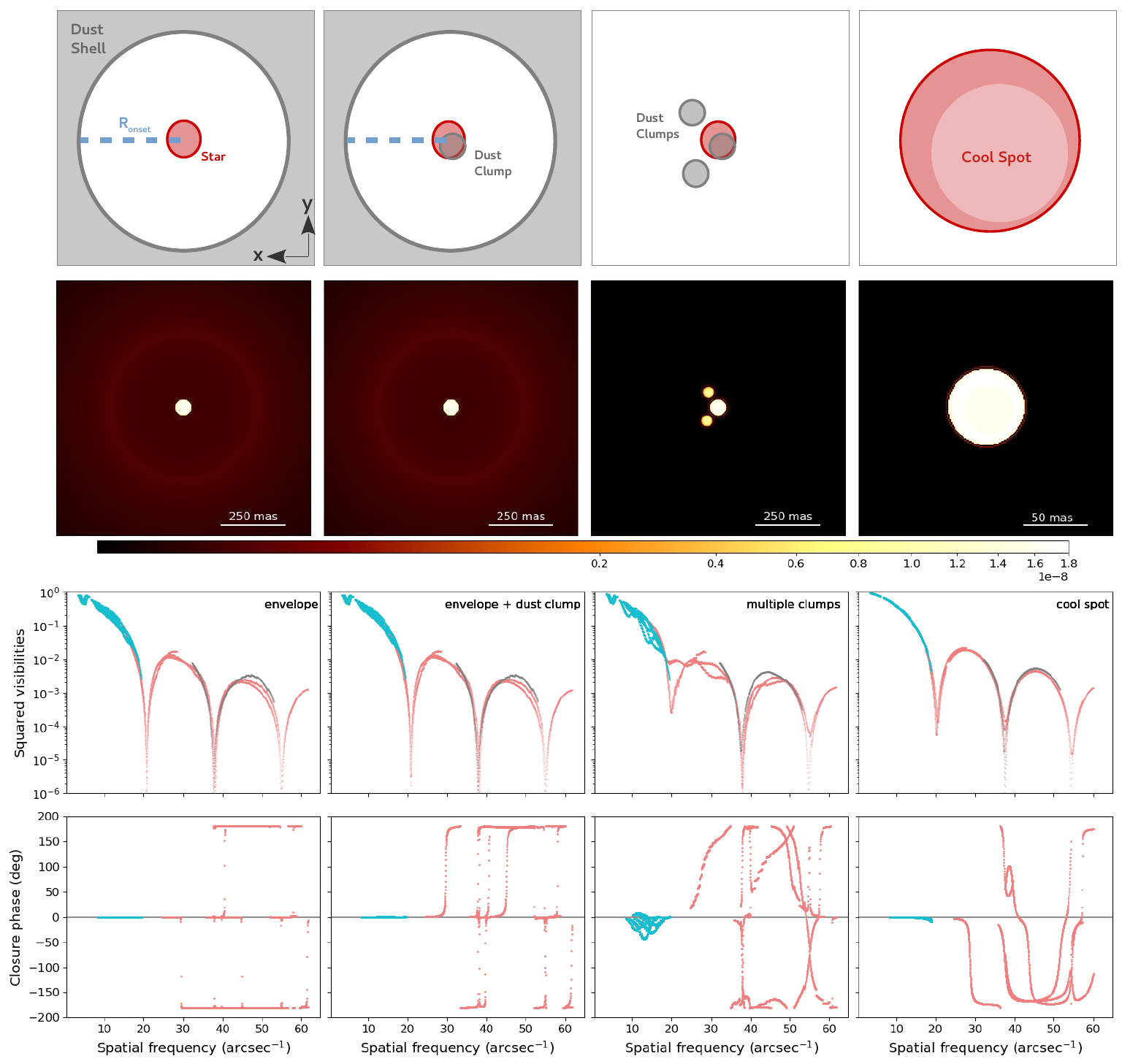}
      \caption{First row: Schematics of the different model setups (not to scale). Second row: Intensity images (W m$^{-2}$ $\mu$m$^{-1}$ arcsec$^{-2}$) output from \textsc{Radmc3D} at 10 $\mu$m for the three dust setups and one for the composite \textsc{Phoenix} models representing the cool spot. The intensity is scaled with a power-law with exponent 0.3 to best show the contribution of the different model elements. The image depicting the cool spot has been zoomed in to better show the stellar disk (see legend for spatial scale used). Third row: Squared visibilities from each of the models in small (blue), medium (pink) and large (grey) configurations. Bottom row: Closure phase for each of the models.
              }
         \label{cp_model}
   \end{figure*}
   
\section{Radiative transfer modelling}
\label{sec:rad_trans}
In order to model the circumstellar dust around Betelgeuse, we use the 3D radiative transfer code \textsc{Radmc3D} (version 2.0, \citealt{2012ascl.soft02015D}). This code enables us to input an arbitrary 3D dust density distribution, dust grain properties and stellar parameters to create cubes of intensity maps spanning the wavelength range of our observations. The interferometric observables, visibilities and closure phases, are then obtained by running these models through a VLTI simulator, \textsc{Aspro2} \citep{2013ascl.soft10005B}. From here we can directly compare our observations and models.  

\subsection{Parameters and assumptions}
\subsubsection{Stellar parameters} 
\label{sec:stellarpar}
\cite{2005ApJ...628..973L} measured a pre-dimming effective temperature, T$_\text{eff}$, of 3650~$\pm$~25\,K, during the dimming this temperature was measured at 3600~$\pm$~25\,K \citep{2020ApJ...891L..37L}, though a subsequent study by \cite{2021RNAAS...5....8Z} suggest a larger temperature drop. As a drop in temperature during the dimming is expected to be localised, (see e.g. \cite{2021Natur.594..365M}), we opted for the higher pre-dimming temperature for our models. We note that a difference of 100~K has limited impact on the continuum SED at these MIR wavelengths. The spectral energy distribution of Betelgeuse was approximated with a \textsc{Phoenix} stellar atmosphere model from \cite{2013A&A...553A...6H} with T$_\text{eff} = 3700$\,K  and surface gravity $\log g = 0$ \citep{2019A&A...627A.138A} scaled by the angular diameter in the K-band (close to the H$^{-}$ opacity minimum), $\theta_\star^\mathrm{NIR}$ = 42.11~mas. However these atmosphere models do not extend to our wavelength range and we therefore extrapolate using the Rayleigh–Jeans law. To test the validity of this atmosphere model and its extrapolation we ran a comparison test with a \textsc{MARCS} stellar atmosphere model from \cite{2000PhDT........16D} with the following parameters: T$_\text{eff}$ = 3600 K, log g = 0.00, [Fe/H] = 0.00 and M= 15 M$_\odot$. We find that the visibilities produced using both models are virtually indistinguishable. Which is unsurprising as the visibilities depend on the relative weights of the components and not the absolute values. We therefore opt to continue with the extrapolated \textsc{Phoenix} model to remain consistent in comparing our findings to \citet{2021Natur.594..365M}. The stellar diameter for the light emitting surface in the N-band was set to $\theta_\star^\mathrm{MIR}$ = 59.02~mas (see section \ref{sec:vis}) at a distance of $222^{+48}_{-34}$\,pc \citep{2017AJ....154...11H,2020ApJ...902...63J}. The MIR-disk diameter is thus $13.1^{+2.9}_{-2.0}$\,au. It is currently not possible to include an inhomogeneous photosphere into the \textsc{Radmc3D} dust models. In section \ref{sec:cool_spot} we test the effect of a cool spot on the photosphere on our observables without the inclusion of dust.

\subsubsection{Dust composition and grain size}
\label{sec:dustcomp}
Led by the findings of \cite{2009A&A...498..127V}, we investigated three dust compositions; olivine (MgFeSiO$_{4}$), alumina (Al$_{2}$O$_{3}$), and the dust mixture found by these authors to best fit the SED of Betelgeuse, comprising of majorly melilite (Ca$_{2}$Al$_{2}$SiO$_{7}$) with smaller amounts of alumina and olivine (mass fractions of 0.64, 0.20 and 0.16 respectively). Optical constants were obtained from \cite{1995A&A...300..503D}, \cite{1997ApJ...476..199B} and \cite{1998A&A...333..188M} for olivine, alumina and melilite respectively. It was clear from early tests that alumina as the sole composition of the radially outflowing wind did not fit the shape of the observed squared visibilities, in particular it does not reproduce the dip at $10 \mu$m that can be seen most prominently at the two shortest baselines (see fig. \ref{vis_comp_AL}). With this in mind, we continued on with the remaining two compositions for the rest of the study (see Sect.~\ref{sec:comp_mdot}). We adopt a dust grain size range of $a = 0.01 - 1~\mu$m from \cite{2009A&A...498..127V} for our spherical grains, with a size distribution $n(a) \propto a^{-3.5}$ \citep{1977ApJ...217..425M} as we deem this more physical than a single grain size in the ambient medium of RSGs where dust nucleation and growth are ongoing. Employing a size distribution also avoids non-physical resonances in the radiative transfer modelling.

      \begin{figure*}[t!]
            \includegraphics[width=\hsize]{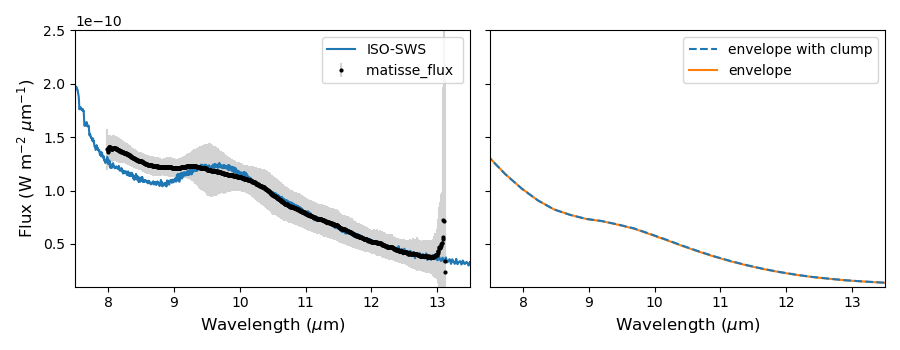}
      \caption{Left: Spectral energy distribution from ISO-SWS01 taken pre dimming (1997-10-08, see Sect. \ref{sec:obs} for details) shown in blue and MATISSE shown in black (error bars shown in grey). Right: Comparison of \textsc{RADMC3D} dust shell models, one with a clump positioned in front of the star and one without. 
              }
         \label{SED}
   \end{figure*}
\subsection{Dust distribution}
\label{sec:morph}

Given the constraints of our data set, we do not aim to reconstruct the exact complex morphology of the inner wind of Betelgeuse. Instead we opt for three relatively simple morphological setups. The idea behind these setups is that they capture the basic components thought to be present in the ambient medium of RSGs. These are that a (roughly) spherical outflow is present through which mass is lost, however, part of the mass loss may be attributed to dusty clumps embedded in this outflow suggesting episodic events of localised mass loss \cite[e.g.,][]{2011A&A...531A.117K,2014A&A...568A..17O}.

With this in mind, the first setup consists of a radial outflow expanding at a constant velocity. Adopting a constant dust-to-gas ratio, the dust density distribution falls off as $r^{-2}$. We take the inner radius of our dust shell to be 13 $R_\star^\mathrm{NIR}$ (see Sect. \ref{sec:vis}), as determined through SED fitting by \cite{2009A&A...498..127V} and a constant expansion velocity, v$_{\text{exp}}$, of 14 km s$^{-1}$ \citep{2010A&A...523A..18D}, where we assume the gas and dust are well mixed and part of one flow, expanding at the same velocity. We then vary both the composition and the dust mass-loss rate (see Sect. \ref{sec:comp_mdot}). \cite{2011A&A...531A.117K} presented VLT/VISIR MIR images of the circumstellar environment of Betelgeuse and identify a partial dust shell at an onset radius of $\approx$ 24 $R_\star^\mathrm{NIR}$, however, these images also show the presence of dust closer in to the star than said shell. For this reason we choose this closer dust shell onset in order to approximate the dusty environment as a whole. For this and the following setup we use a spherical grid (n$_r$, n$_\theta$, n$_\phi$) = (20, 20, 20), spanning radii from 8 to 400 au (1.7 - 86 $R_\star^\mathrm{NIR}$), and employ four levels of adaptive mesh refinement on top of this grid to adequately resolve the dusty regions.

The second setup adds a dusty clump positioned in front of the star in addition to the dusty envelope. The size, position and density of the homogeneous and spherical clump are taken from \cite{2021Natur.594..365M}, where they were determined through comparison of radiative transfer models to images from VLT/SPHERE-ZIMPOL. Our clump parameters are therefore; $r_{\rm c} =$ 4.5 au, $x_{\rm c}$ = -1.9 au, $y_{\rm c}$ = -1.8 au, $z_{\rm c}$ = 20 au (where $x$ and $y$ are indicated in Fig. \ref{cp_model} and $z$ is line of sight position which is positive toward the observer) and the clump dust density is $\rho_{\rm c} = 2.0 \times 10^{-18}$ g cm$^{-3}$.

The third dust morphology setup we investigated removes the dust shell and places two more clumps in the close environment around the star, with identical properties to that described above. For this test we use olivine as the dust composition for all clumps. The decision to add two clumps for this test scenario was motivated by previous SPHERE/ZIMPOL observations of the dust in the inner environment of the RSGs Betelgeuse and Antares, which both show a few patches of larger dust density within several stellar radii (\citealt{2016A&A...585A..28K}, \citealt{2021MNRAS.502..369C}). We point out that our aim is not to try to constrain the properties and spatial configuration of an inner clumpy environment -- apart from modelling challenges, we lack the ($u,v$) coverage to attempt this, rather to probe the effects of this type of morphology on the visibility and closure phase signal. Schematics of all three \textsc{Radmc3D} models are shown in Fig. \ref{cp_model}.

\subsection{Modelling results}
\label{sec:mod_results}

\subsubsection{First setup: composition and mass-loss rate}
\label{sec:comp_mdot}

After a coarse exploration of parameter space to map the effect of the dust mass-loss rate on the visibilities, we zoomed in on the range from  $1 \times 10^{-10}\,$ to  $1 \times 10^{-9}\,$~\msunyr\ in steps of $1 \times 10^{-10}\,$~\msunyr for our two main chemical compositions. To assess how well our models recreate the observations we perform a $\chi^2$ estimation, simultaneously on both small configuration snapshots. 
We give an estimate of the confidence interval by normalising such that the best model has a reduced $\chi^2$ ($\chi^2_{\nu}$) value of 1, from here we calculate the P-value. All models with a P-value higher than 0.05 (within the 95\% confidence interval) are deemed acceptable models from which we take our parameter ranges. To account for our limited sampling rate we perform a linear interpolation to the upper P-value points to give a better estimate on the confidence interval of the dust mass-loss rate. Considering the coarse sampling of our model grid, due to the exploratory nature of the dust geometries, using other statistical techniques is not possible. Since the main uncertainty lies in the model geometry and our (u, v) coverage, we do not deem necessary to expand our model grid. The current estimates are meant to give an order of magnitude of the dust mass-loss rate from the MATISSE observations.

Here, we find that both an outflow composed of just olivine and one with the mixture of melilite, alumina and olivine, give us similarly good matches (see Fig. \ref{vis_comp}). 
We find dust mass-loss rates of $3^{+0.9}_{-0.9}~\times 10^{-10}\,$~\msunyr and $4^{+0.9}_{-0.9}~\times~10^{-10}\,$~\msunyr for olivine and the melilite mix respectively. However, it is worth noting that a change of $R_{\rm in}$ or $v_{\rm exp}$, which were kept constant in this study, would have an impact on the derived dust mass-loss rate. Since the model input here assumes a spherically symmetric outflow we do not expect to see the closure phases depart from 0 or $\pm$ 180$^{\circ}$ in line with the model shown in Fig. \ref{cp_model}. The few stray points in the closure phases are a result of the minor asymmetries caused by the pixel size (3.6 mas) of our model images inputted to \textsc{Aspro2}.

\subsubsection{Second setup: impact of the dusty clump}
\label{sec:clump_impact}
The addition or omission of the dusty clump in front of the star, described in Sect. \ref{sec:morph}, appears to have little effect on the visibility shape giving almost the exact same $\chi^2_{\nu}$ value either way. This is also evident when comparing the spectral energy distributions of our models: the addition of the dusty clump in the line of sight, makes a negligible impact (see Fig.~\ref{SED}).  We return to this in Sect. \ref{sec:discussion}. 
In line with these findings are the model images at 10\,$\mu$m in Fig.~\ref{cp_model}, second row, where both the images with and without clump are virtually indistinguishable. Where we do clearly see the signature of the clump is in the closure phases of the medium configuration, as seen in the fourth row of Fig. \ref{cp_model}. Here we can see departures from symmetry of our model at the same spatial scales as the observations, however, the observed closure phase (Fig. \ref{observations}) signal is much more complex and unlikely the result of a singular dusty clump. We note that the temperature of the dust in this clump ranges from approximately 1200\,K - 1800\,K. Only a minor fraction (less than 10\%) has a temperature in excess of 1500\,K.

\subsubsection{Third setup: multi-clump model}
\label{sec:multi_clump}
\begin{figure*}[]
            \includegraphics[width=\hsize]{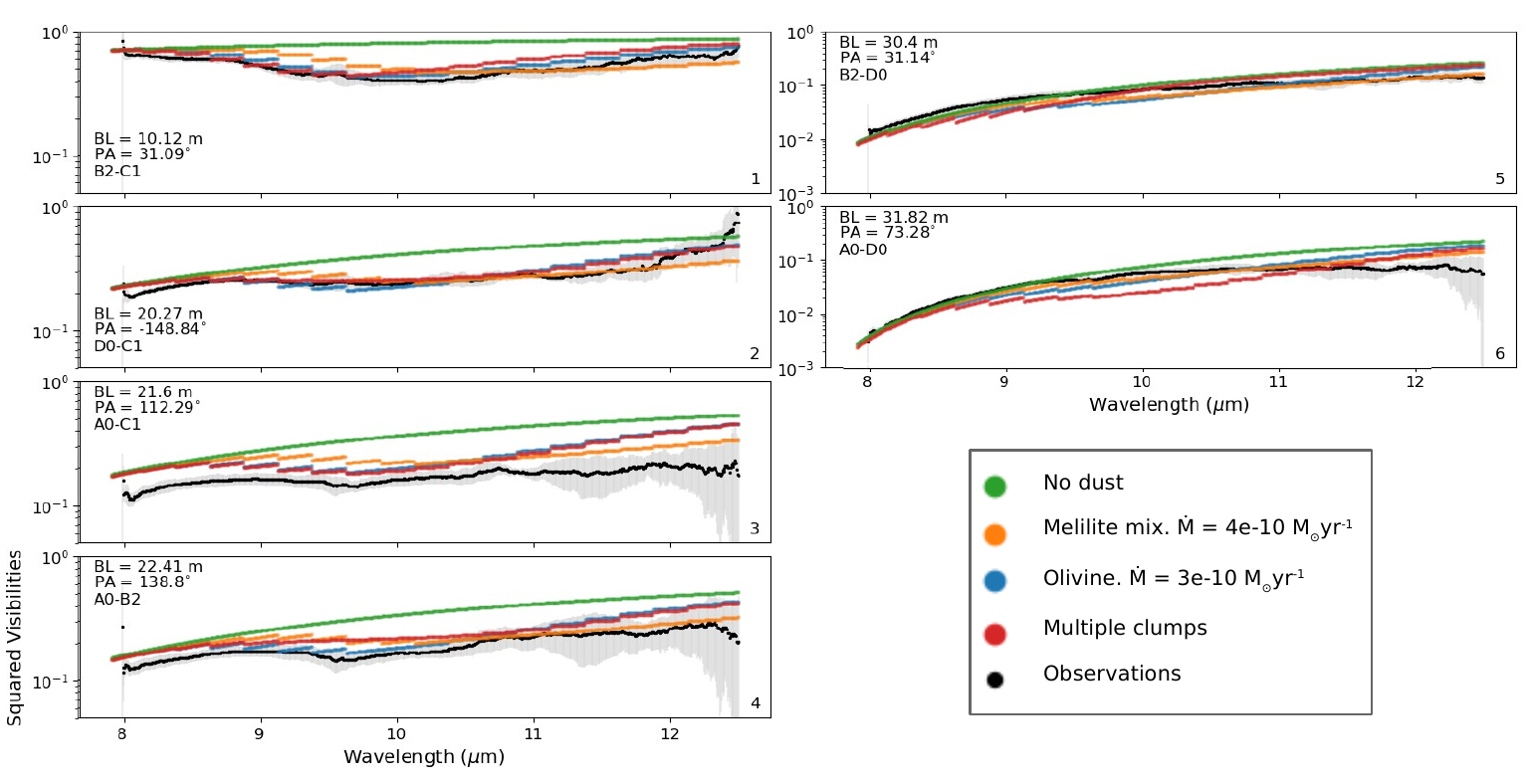}
      \caption{Comparison of visibilities from \textsc{RADMC3D} models to the small configuration observations (2020-02-08(C)) split up by baseline. The same models compared to 2020-02-08(A) are shown in Fig. \ref{vis_comp2} and the residuals are plotted in Figs. \ref{res_p3} and \ref{res_p1}. BL and PA denote the baseline length and position angle respectively. For $\chi^2_{\nu}$ values see Table \ref{chi}.
           }
         \label{vis_comp}
   \end{figure*}

The purpose behind the multi-clump model is not to reproduce the exact dust morphology in the inner environment of Betelgeuse but to evaluate what effect multiple dust clumps would have on the visibilities and closure phases. A parameter study of possible clump configurations with the aim to match the closure phase signal is currently far beyond what is technically possible with our 3D models given the large parameter space, and the poor constraints provided by the ($u, v$) coverage of our data.
For the multi-clump demonstration setup we kept the dust clump in the line of sight of the star at a fixed position. We then placed two more identical dust clumps in positions within 20 au of the star (but in the plane of the sky, i.e. at $z=0$). In order to find a tentative match for our visibilities, we then rotated the clumps around the star. As is apparent in the second row of Fig. \ref{cp_model}, these additional clumps are much more visible at 10\,$\mu$m than the clump in front of the star. While we did not aim to fit the visibilities with this simple test we obtain a reasonable match to most baselines, shown in Figs.~\ref{vis_comp} and \ref{vis_comp2}, and Table.~\ref{chi}. This multiple clump model also creates more complex closure phases (see Fig. \ref{cp_model}) which seem to better mimic the closure phase structure as seen in the observations.

\begin{table}[]
\caption{Comparison between the observed and modelled squared visibilities showing $\chi^2_{\nu}$ values for each model shown in Fig. \ref{vis_comp} and \ref{vis_comp2}. 
}
\label{chi}
\centering          
\begin{tabular}{l l }       
\hline\hline       
                
Model & $\chi^2_{\nu}$  \\ 
\hline  \noalign{\vskip 2mm}    
    Envelope - Melilite mix  & 3.02 \\
    Envelope - Olivine  & 3.58 \\
    Multiple clumps & 7.96\\
    No dust  & 20.31\\

   \noalign{\vskip 2mm}  

\hline                 
\end{tabular}
\end{table}


\section{Cool spot model}
\label{sec:cool_spot}
In order to test if the MATISSE observations are consistent with a cool spot on the surface we create a composite model following \citet{2021Natur.594..365M}. We take the area of the stellar photosphere to be $\pi(59.02/2)^2$~mas$^2$ and assign an effective temperature of 3700\,K (again scaling the stellar atmosphere models by $\theta_\star^\mathrm{NIR}$). We then define a circular cool patch of radius 19~mas and centre ($x,y$) at (-2.4,-2.4) mas with respect to the stellar centre, on the surface of the star with a temperature of 3400\,K. The composite image is shown in row 2 of Fig.~\ref{cp_model}, where it can be seen that the cool patch on the surface does not cause as high of a brightness contrast at 10\,$\mu$m as it does in the visible: at 10\,$\mu$m the specific intensity is 9.4\% less in the cool spot than elsewhere on the stellar surface.

As predicted, the model visibilities now lack the 10\,$\mu$m feature seen in the observations. However, the addition of a dusty envelope as described in Sect.~\ref{sec:comp_mdot}, or other shapes of circumstellar dust, to the cool spot model would remedy this. The cool patch is clearly seen in the closure phases, showing significant departure from symmetry; see Fig.~\ref{cp_model}.

\section{Discussion}
\label{sec:discussion}

\subsection{Mass-loss rate and gas-to-dust ratio}

From our spherical wind models we obtained dust mass-loss rates in the range of (2.1 - 4.9)~$\times~10^{-10}\,$~\msunyr which agree well with those determined by \cite{2009A&A...498..127V}. \cite{2010A&A...523A..18D} derive a total mass-loss rate for Betelgeuse of 2.1~$\times 10^{-7}\,$~\msunyr from CO line profiles, however, \cite{1994ApJ...424L.127H} show that CO is under-associated in the circumstellar envelope of Betelgeuse suggesting this rate could be a lower limit. Hence, combining this mass-loss rate with our models this suggests a minimum gas-to-dust ratio between 430 and 1000 for a homogeneous wind. A similarly high gas-to-dust ratio for Betelgeuse ($\geq$ 550) and for the RSG Antares ($\geq$ 600) was found by \cite{1999A&A...345..605J}. This suggests the dust formation in the outflow is not particularly efficient. It is possibly only efficient in clumps.

\subsection{Detecting dust potentially formed during the Great Dimming}

\begin{figure}[t!]
   \centering
   \includegraphics[width=\hsize]{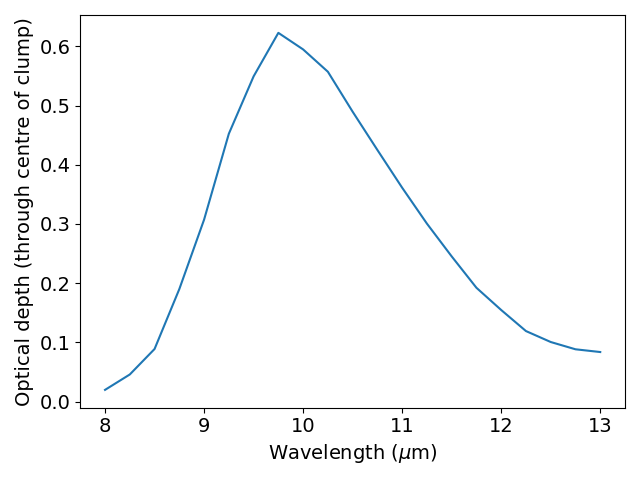}
      \caption{The optical depth through the centre of the dust clump, composed of olivine, as a function of wavelength as computed by \textsc{Radmc3D}. Beware that the clump is not covering the entire star.
              }
         \label{tau}
   \end{figure}

\begin{figure}[t!]
   \centering
   \includegraphics[width=\hsize]{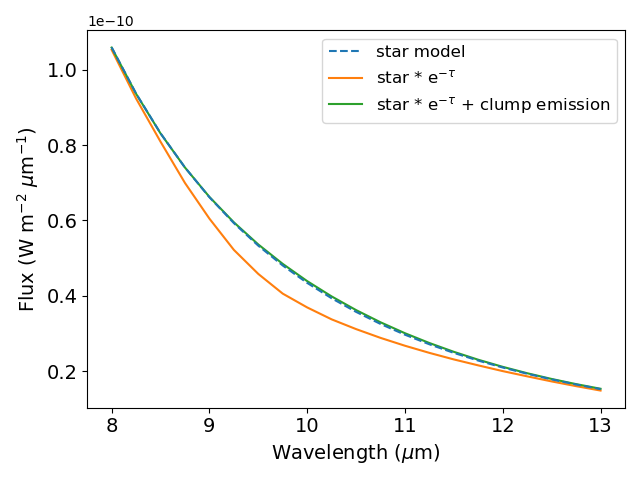}
      \caption{The blue dashed line shows the SED (computed from the \textsc{Radmc3D} images outputted at different wavelengths) of the star with no dust. The orange line shows the SED of the star after each image is multiplied by the transmittance map, T $= e^{-\tau}$ where $\tau$ is the optical depth, caused by the dust clump in the line of sight. Adding the dust clump emission to these new images results in the green line, which happens to almost overlap with the blue dashed line.
              }
         \label{T_emission}
   \end{figure}

The addition of the singular clump in front of the stellar disk with similar properties as derived in \citet{2021Natur.594..365M} does not significantly change the visibility signal. The dust clump also has a negligible contribution to the SED (right panel Fig.~\ref{SED}) at these relatively long wavelengths in line with SED observations prior and during the dimming event (left panel Fig.~\ref{SED}). Further investigation shows that the optical depth of the dust clump is such that \textit{the extinction (including scattering and absorption) caused by the clump almost exactly compensates for the dust emission}, see Figs.~\ref{tau} and \ref{T_emission}. This is not the case, however, if we use a different set of clump parameters, such as a clump with a higher density for example. In which case, the NIR flux in the SED is reduced by the clumps presence in the line of sight. We note that if this clump was not in the line of sight, such as the other two clumps in our multi-clump setup, the IR emission does impact the SED.

\cite{2020ApJ...897L...9D} use a spherical shell around the star to model the dust that would have caused the dimming showing that the extra dust emission should have been detected in the SED at these wavelengths, using this as part reason to exclude dust as a possible cause of the `Great Dimming'. \cite{2020ApJ...905...34H} use similar reasoning, a lack of increase in emission compared to previous infrared photometry, as an argument against the dust hypothesis. Our modelling, however, shows that the localised presence of the newly formed dust in the line of sight of the star would go undetected at these wavelengths using these observational methods. 

\citet{2021Alexeeva} concluded from TiO line fitting to spectroscopic observations in the NIR at wavelengths shorter than 1~$\mu$m, that only a cool photospheric patch could reproduce their observations. \citet{2021Natur.594..365M}, however, found that a cool spot model produces a reduction of the flux at 1.6 $\mu$m that is significantly larger than what is observed during the dimming event. Further, \cite{2022NatAs.tmp..121T} detect an enhancement in the optical depth at 10 $\mu$m during the dimming event, indicating the new formation of dust in the line of sight of the star.

\subsection{Clump or clumps?}

While we do not fit the closure phases, it is clear from both the dust modelling (Sect.~\ref{sec:rad_trans}) and cool spot modelling (Sect.~\ref{sec:cool_spot}), that dust clumps (patches with local density enhancements) and surface features are the keys to reproduce the complex features seen in the MATISSE observations. This is unsurprising given existing observations and theory of RSGs surfaces and winds that point to variations on spatial scales comparable to or smaller than the stellar diameter (see Sect.~\ref{sec:intro}). While the addition of the single clump of dust, described in Sect. \ref{sec:morph}, in the line of sight of the star does cause some signal in the closure phase it cannot fully explain the complexity of the observed closure phases. The situation is somewhat improved for our cool spot model. Therefore, we can not at present conclusively rule out either of these scenarios or indeed the presence of both simultaneously as concluded by \citet{2021Natur.594..365M}.

Our pilot `multiple clump' scenario shows that one may reproduce the visibilities and produce more intricate closure phase signals using clumps of dust in the inner circumstellar environment. While this model aims to show a proof of concept ample evidence supports the presence of clumps, therefore it seems more physically realistic than our other configurations including the outflow models starting at 13 R$_\star^\mathrm{NIR}$. The presence of clumps is supported by previous spatially resolved observations. One such observation by \cite{2016A&A...585A..28K} using VLT/SPHERE/ZIMPOL shows a clumpy polarisation signal within 3 R$_\star$ which indicates the presence of a patchy dusty environment. The patchy nature of the inner environment appears to show similarities to further out regions in the wind, with observations with VLT/VISIR by \cite{2011A&A...531A.117K} identifying large dust clumps in the outer wind.   Betelgeuse is not the only RSG to show these clumpy features. Also for Antares, whose clumpy wind has also been observed with VLT/SPHERE/ZIMPOL, VLT/VISIR and MIRLIN at KECK II (\citealt{2021MNRAS.502..369C}, \citealt{2014A&A...568A..17O} and \citealt{2001ApJ...548..861M}, respectively) these inhomogeneities have been detected. NOEMA observations of $\mu$ Cep show that mass lost through clumps accounts for $\geq$ 25\% of the RSG stars total mass-loss \citep{2019MNRAS.485.2417M}. \cite{2021AJ....161...98H} also find dust clumps around the RSG VY CMa. Using data from the Hubble Space Telescope, they calculate the outward motion of these clumps allowing their ejection time to be calculated. The authors find that the ejection times of some of these clumps correspond with minima in the light curve of the star.


\section{Conclusions}
\label{sec:summary}

We obtained MATISSE N-band observations in February 2020 during the `Great Dimming' of Betelgeuse such as to probe the thermal emission of dust in the immediate surroundings of the star. Ideally, such observations may help to distinguish between the hypotheses proposed to explain this extraordinary decrease in visual brightness. To this end, we modelled the visibilities observed in the small VLTI configuration and examined the closure phases in the small and medium configurations.

From our parametric modelling of the VLTI/MATISSE data, we determined the uniform disk diameter of the star between 8 and 8.75 $\mu$m to be 59.02 $\pm$ 0.64 mas or a radius 1409$^{+319}_{-229}$\,R$_{\odot}$ at the adopted distance of $222^{+48}_{-34}$\,pc. With the stellar size constrained, our three adopted dust models match the visibility data well: a spherical wind, a spherical wind with a dust clump in the line of sight, and three clumps placed around the star of which one is crossing the line of sight. This implies that these visibility data are not sufficiently sensitive to the spatial distribution and composition of dust in the field of view of MATISSE to accurately map the dust in the star's environment. Better ($u,v$) coverage is needed to obtain a more complete image of the dust spatial morphology. Fitting a spherical homogeneous wind to the visibilities gives dust mass-loss rates of (2.1 - 4.9) $\times 10^{-10}$~\msunyr suggesting a minimum gas-to-dust ratio between 430 and 1000 compared to the gas mass loss estimate by \citet{2010A&A...523A..18D}, contrary to much lower values for outflows of Asymptotic Giant Branch stars (where the canonical value is 100-200). This high gas-to-dust ratio indicates that dust formation may not be efficient in the wind or perhaps only efficient in clumps. Our models also exclude the possibility that the wind is dominated by Al$_2$O$_3$. However, this does not rule out that Al$_2$O$_3$ could be dominant in dust clumps in the inner wind (see e.g. \citealt{2007A&A...474..599P,2021MNRAS.502..369C}). The complexity of the closure phases from MATISSE suggests major asymmetries in the field of view ($\sim$ 1 arcsec), these could be caused by an asymmetric stellar disk, large scale surface features or dust clumps in the wind, possibly a combination of all three. While our current data set in principle allows for a homogeneous wind (to reproduce visibilities) with stellar surface features (to account for the non-zero closure phases) there is observational evidence from VLT/SPHERE \citep{2016A&A...585A..28K} and VLT/VISIR \citep{2011A&A...531A.117K} that clearly show an inhomogeneous dust distribution around the star. 

In terms of the `Great Dimming' of Betelgeuse we found that both models, cool spot and dust clump, or a combination of these models, are compatible with the observations. In particular, we find no inconsistencies with the modelling results presented by \cite{2021Natur.594..365M}.  We note that the dust clump from \citeauthor{2021Natur.594..365M}, positioned in the line of sight of the star, would be undetectable in the SED and visibilities in the mid-infrared as the extinction caused by the dust is directly compensated for by the dust emission. From our models, we can see that the presence of such a clump would only be detectable in the closure phases from the medium configuration (which probes scales of 20~-~80~mas). This clearly shows that in order to understand the nature of the `Great Dimming', high angular resolution is mandatory to distinguish the photosphere from the circumstellar environment. 

Future VLTI/MATISSE observations of red supergiants and cool evolved stars, with good ($u,v$) coverage, will allow us to observe this close region where dust nucleation takes place in more detail. Mapping the dust across multiple epochs combined with ALMA observations of the gas morphology and kinematics would provide further insight into the connection of the geometry in the dust forming region and the mass-loss properties and mechanism(s) of RSGs.


\begin{acknowledgements}
The authors acknowledge funding from the KU Leuven C1 grant MAESTRO C16/17/007. J.S.B. acknowledges the support received from the UNAM PAPIIT project IA 101220 and from the CONACYT project CF-2019-263975.  
This research has benefited from the help of SUV, the VLTI user support service of the Jean-Marie Mariotti Center \footnote{\url{http://www.jmmc.fr/suv.htm}}.
This research has made use of the Jean-Marie Mariotti Center \texttt{Aspro} service \footnote{Available at \url{http://www.jmmc.fr/aspro}}. This research has also made use of the Jean-Marie Mariotti Center JSDC catalogue\footnote{available at \url{http://www.jmmc.fr/catalogue_jsdc.htm}}.

\end{acknowledgements}

\bibliographystyle{aa}
\bibliography{references}

\begin{appendix}
\section{Additional comparison plots of \textsc{RADMC3D} models and MATISSE visibilities}
\label{sec:appA}

      \begin{figure}[ht!]
            \includegraphics[width=\textwidth]{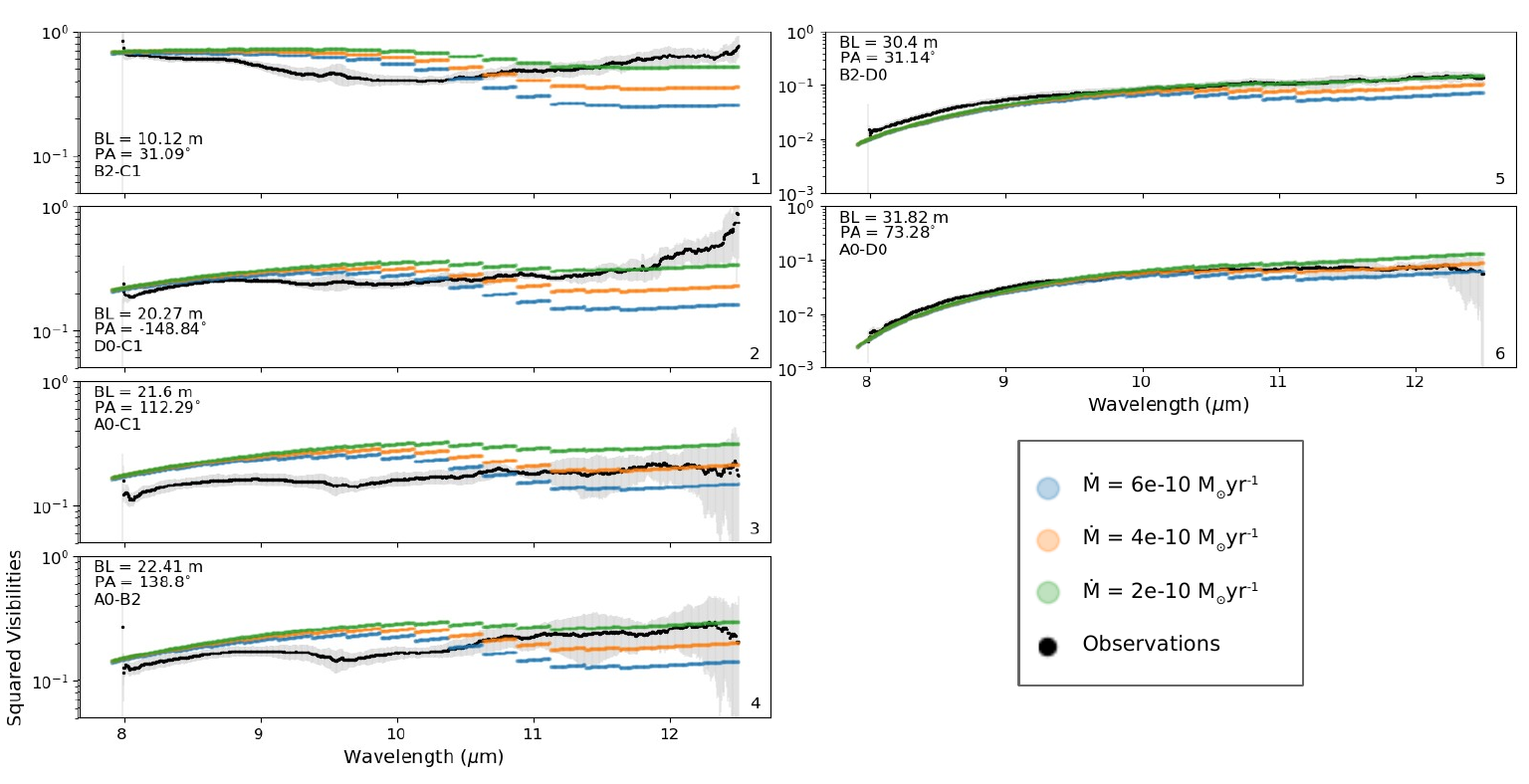}
      \caption{Comparison of visibilities from \textsc{RADMC3D} of a radial outflow represented by a spherical dust shell composed of Alumina to the small configuration observations (2020-02-08(C)) split up by baseline. BL and PA denote the baseline length and position angle respectively.
            }
         \label{vis_comp_AL}
   \end{figure}

      \begin{figure}
            \includegraphics[width=\textwidth]{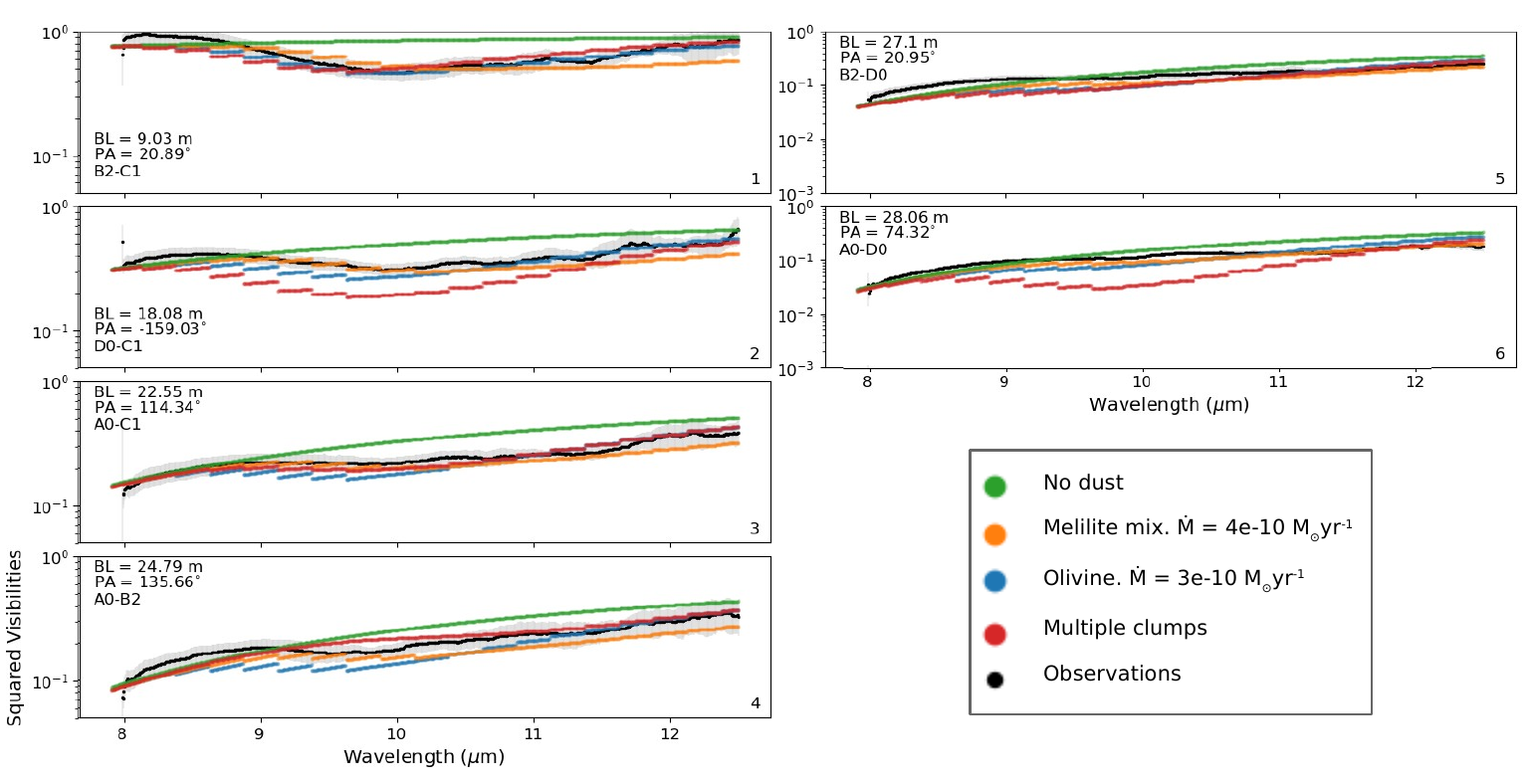}
      \caption{Same as Fig. \ref{vis_comp} but for snapshot 2020-02-08(A)
            }
         \label{vis_comp2}
   \end{figure}

      \begin{figure*}
            \includegraphics[width=\textwidth]{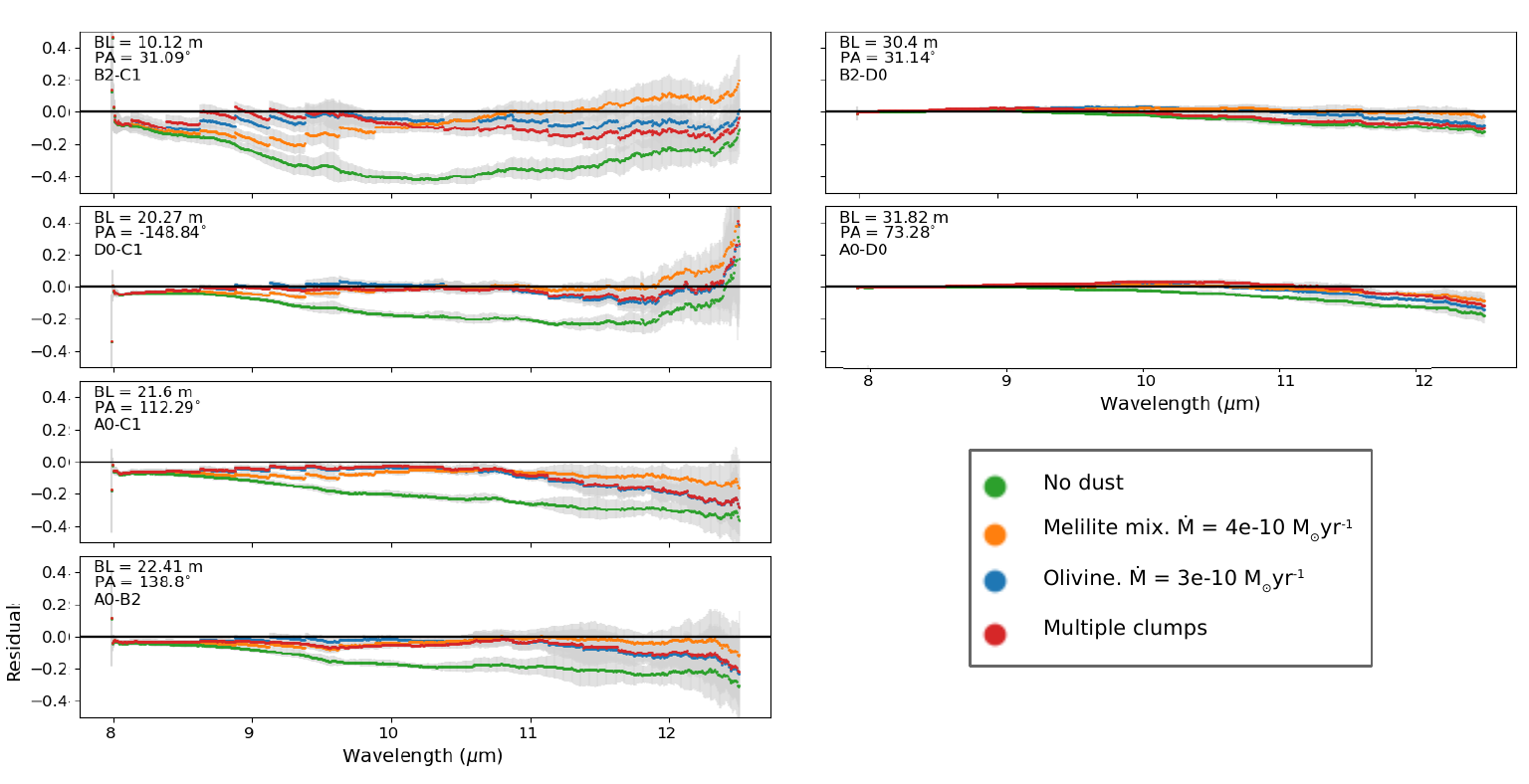}
      \caption{Residual of the squared visibilities of the small configuration (2020-02-08(C)) with respect to the RADMC3D models.
            }
         \label{res_p3}
   \end{figure*}
   
         \begin{figure*}
            \includegraphics[width=\textwidth]{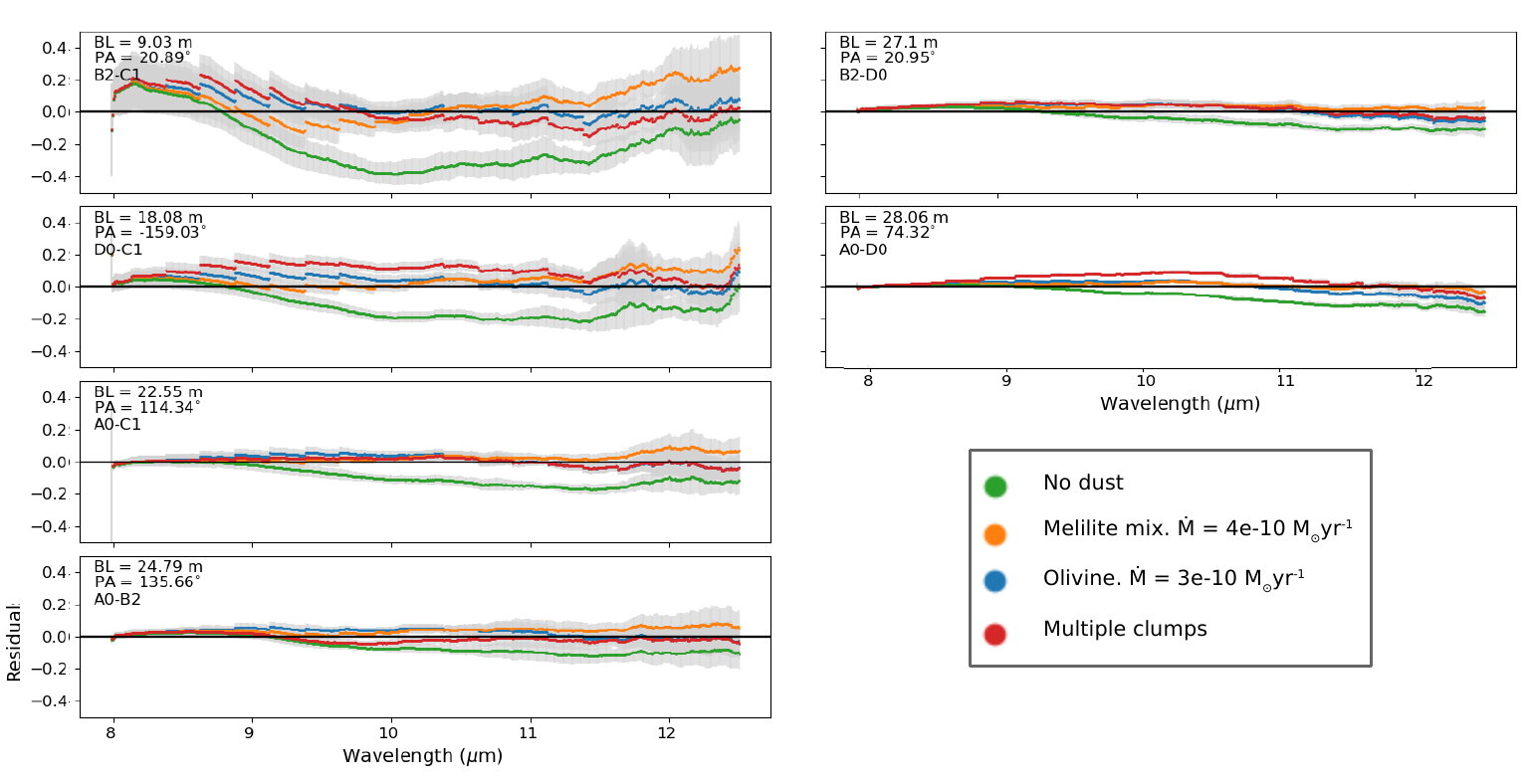}
      \caption{ Same as Fig. \ref{res_p3} but for snapshot 2020-02-08(A).
            }
         \label{res_p1}
   \end{figure*}

\end{appendix}

\end{document}